\pdfoutput=1

\documentclass[11pt]{article}

\usepackage{acl}

\usepackage{times}
\usepackage{latexsym}

\usepackage[T1]{fontenc}

\usepackage[utf8]{inputenc}

\usepackage{microtype}

\usepackage{bbding}
\usepackage{bm}
\usepackage{amsmath}
\usepackage{algorithm}
\usepackage{algorithmic}
\usepackage{amsmath,graphicx,subfigure}
\usepackage{amssymb}
\usepackage{multirow}
\usepackage{booktabs}
\usepackage{arydshln}
\usepackage{hyperref}
\usepackage{color,soul}
\usepackage{natbib}
\usepackage{xurl}

\title{Cross-Utterance Conditioned VAE for Non-Autoregressive Text-to-Speech}

\author{
Yang Li\textsuperscript{1,}\footnotemark[1] , 
Cheng Yu\textsuperscript{1,}\footnotemark[1] , 
Guangzhi Sun\textsuperscript{2,}\footnotemark[2] , 
Hua Jiang\textsuperscript{3}, 
Fanglei Sun\textsuperscript{1,}\footnotemark[2] , 
Weiqin Zu\textsuperscript{1}, \\
\bf Ying Wen\textsuperscript{4}, 
\bf Yang Yang\textsuperscript{1}, 
\bf Jun Wang\textsuperscript{5} \\ 
\textsuperscript{1}ShanghaiTech University,
\textsuperscript{2}{Cambridge University},
\textsuperscript{3}Neurowave Ai Limited,\\
\textsuperscript{4}{Shanghai Jiao Tong University}
,\textsuperscript{5}University College London}

\begin{document}
\maketitle

\renewcommand{\thefootnote}{\fnsymbol{footnote}}
\footnotetext[1]{These authors contributed equally to this work.}
\footnotetext[2]{Corresponding authors. E-mails: gs534@cam.ac.uk; sunfl@shanghaitech.edu.cn}

\begin{abstract}
Modelling prosody variation is critical for synthesizing natural and expressive speech in end-to-end text-to-speech (TTS) systems. 
In this paper, a cross-utterance conditional VAE (CUC-VAE) is proposed to estimate a posterior probability distribution of the latent prosody features for each phoneme by conditioning on acoustic features, speaker information, and text features obtained from both past and future sentences. 
At inference time, instead of the standard Gaussian distribution used by VAE, CUC-VAE allows sampling from an utterance-specific prior distribution conditioned on cross-utterance information, which allows the prosody features generated by the TTS system to be related to the context 
and is more similar to how humans naturally produce prosody. The performance of CUC-VAE is evaluated via a qualitative listening test for naturalness, intelligibility and quantitative measurements, including word error rates and the standard deviation of prosody attributes. Experimental results on LJ-Speech and LibriTTS data show that the proposed CUC-VAE TTS system improves naturalness and prosody diversity 
with clear margins. 
\end{abstract}

\section{Introduction}
\label{sec:intro}

Recently, abundant research have been performed on modelling variations other than the input text in synthesized speech such as background noise, speaker information, and prosody, as those directly influence the naturalness and expressiveness of the generated audio. 
Prosody, as the focus of this paper, collectively refers to the stress, intonation, and rhythm in speech, and has been an increasingly popular research aspect in end-to-end TTS systems
~\citep{Oord2016WaveNet, Wang2017PTacotron,Stanton2018styletokens, Elias2021PTacotron, chen2021speechbert}.
Some previous work captured prosody features explicitly using either style tokens or variational autoencoders (VAEs) \citep{vae, hsu2018hierarchical} which encapsulate prosody information into latent representations. Recent work achieved fine-grained prosody modelling and control by extracting prosody features at phoneme or word-level~\citep{lee2019finegrained,sun2020generating,sun2020finegrainedvae}.
However, the VAE-based TTS system lacks control over the latent space where the sampling is performed from a standard Gaussian prior during inference.
Therefore, recent research~\citep{2019Dahmaniconditional, Karanasou2021learned} employed a conditional VAE (CVAE)~\citep{cvae} to synthesize speech from a conditional prior.
Meanwhile, pre-trained language model (LM) such as bidirectional encoder representation for Transformers (BERT)~\citep{devlin2018bert} has also been applied to TTS systems~\citep{ hayashi2019pre,Kenter2020RNNbert,Jia2021PnGBERT,Futamata2021preJapan,cong2021controllable} to estimate prosody attributes implicitly from pre-trained text representations within the utterance or the segment. 
Efforts have been devoted to include cross-utterance information in the input features to improve the prosody modelling of auto-regressive TTS \citep{xu2021improving}.
 
To generate more expressive prosody, while maintaining high fidelity in synthesized speech, a cross-utterance conditional VAE (CUC-VAE) component is proposed, which is integrated into and jointly optimised with FastSpeech 2~\citep{ren2020fastspeech}, a commonly used non-autoregressive end-to-end TTS system. 
Specifically, the CUC-VAE TTS system consists of cross-utterance embedding (CU-embedding) and cross-utterance enhanced CVAE (CU-enhanced CVAE).
The CU-embedding takes BERT sentence embeddings from surrounding utterances as inputs and generates phoneme-level CU-embedding using a multi-head attention~\citep{Vaswani2017Attention} layer where attention weights are derived from the encoder output of each phoneme as well as the speaker information. 
The CU-enhanced CVAE is proposed to improve prosody variation and to address the inconsistency between the standard Gaussian prior, which the VAE-based TTS system is sampled from, and the true prior of speech.
Specifically, the CU-enhanced CVAE is a fine-grained VAE that estimates the posterior of latent prosody features for each phoneme based on acoustic features, cross-utterance embedding, and speaker information. 
It improves the encoder of standard VAE with an utterance-specific prior. 
To match the inference with training, the utterance-specific prior, jointly optimised with the system, is conditioned on the output of CU-embedding. 
Latent prosody features are sampled from the derived utterance-specific prior instead of a standard Gaussian prior during inference.

The proposed CUC-VAE TTS system was evaluated on the LJ-Speech read English data and the LibriTTS English audiobook data. In addition to the sample naturalness measured via subjective listening tests, the intelligibility is measured using word error rate (WER) from an automatic speech recognition (ASR) system, and diversity in prosody was measured by calculating standard deviations of prosody attributes among all generated audio samples of an utterance. Experimental results showed that the system with CUC-VAE achieved a much better prosody diversity while improving both the naturalness and intelligibility compared to the standard FastSpeech 2 baseline and two variants. 

The rest of this paper is organised as follows. Section~\ref{sec: preliminaries} introduces the background and related work. 
Section~\ref{sec: model} illustrates the proposed CUC-VAE TTS system. 
Experimental setup and results are shown in Section~\ref{sec:exp_setup} and Section~\ref{sec:exp_results}, with conclusions in Section~\ref{sec: con}.

\begin{figure*}
\centering
    \centering
    \includegraphics[width=\textwidth]{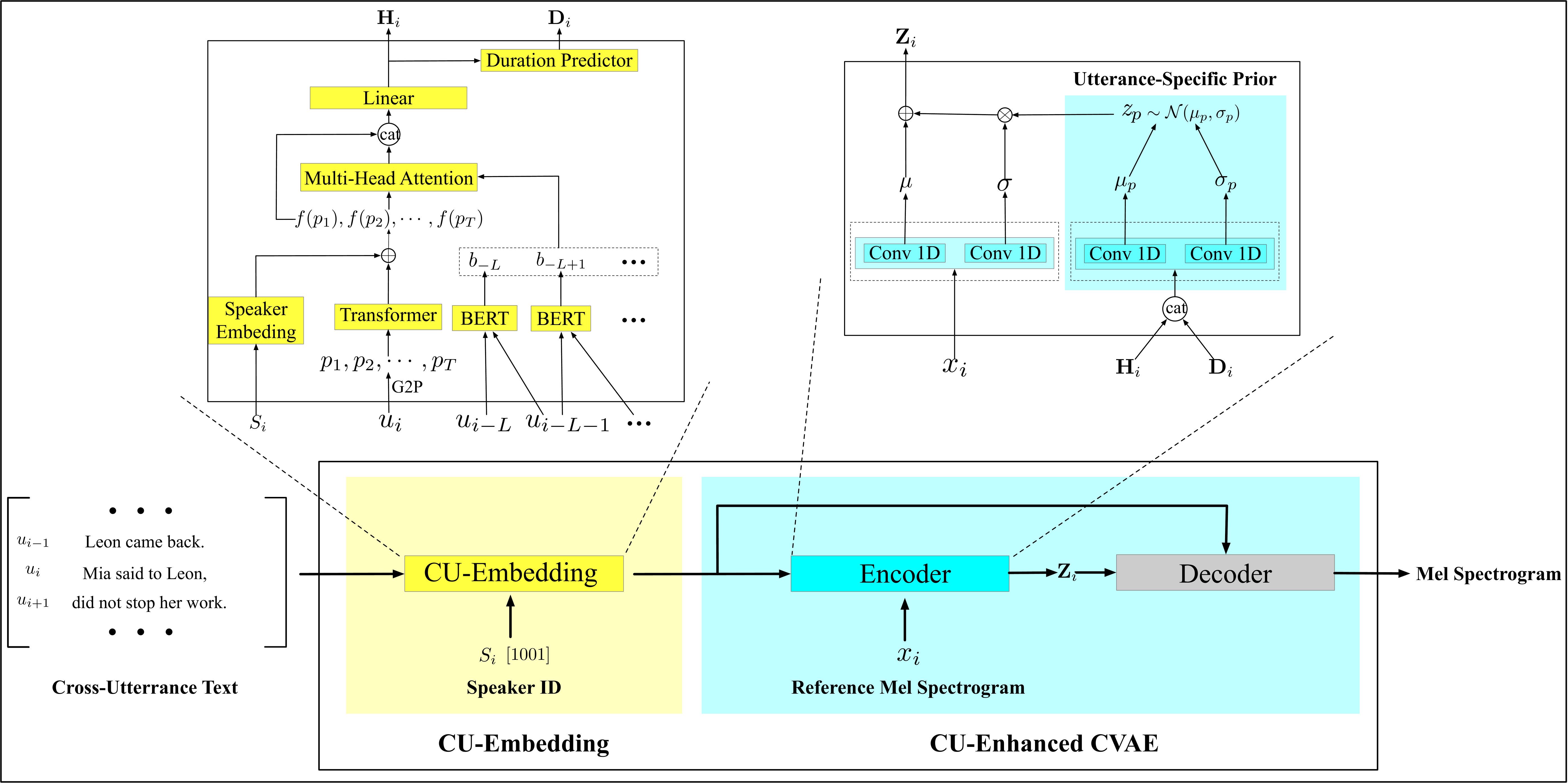}
    \caption{
    The CUC-VAE TTS system architecture consists of the cross-utterance embedding (CU-embedding) and the cross-utterance enhanced (CU-enhanced) CVAE, which are integrated into and jointly optimised with the FastSpeech 2 system.
    }
    \label{fig:model}
\end{figure*}

\section{Background}
\label{sec: preliminaries}
\textbf{Non-Autoregressive TTS.}
Promising progress has taken place in non-autoregressive TTS systems to synthesize audio with high efficiency and high fidelity thanks to the advancement in deep learning. 
A non-autoregressive TTS system maps the input text sequence into an acoustic feature or waveform sequence without using the autoregressive decomposition of output probabilities.
FastSpeech~\citep{ren2019fastspeech} and ParaNet~\citep{peng2019paralltts} requires distillation from an autoregressive model, while more recent non-autoregressive TTS systems, including FastPitch~\citep{Lancucki2021FastPitch}, AlignTTS~\citep{Zeng2020AlignTTS} 
and FastSpeech 2~\citep{ren2020fastspeech}, do not rely on any form of knowledge distillation from a pre-trained TTS system.
In this paper, the proposed CUC-VAE TTS system is based on FastSpeech 2.
FastSpeech 2 replaces the knowledge distillation for the length regulator in FastSpeech with mean-squared error training based on duration labels, which are obtained from frame-to-phoneme alignment to simplify the training process. Additionally, FastSpeech 2 predicts pitch and energy from the encoder output, which is also supervised with pitch contours and L2-norm of signal amplitudes as labels respectively. 
The pitch and energy prediction injects additional prosody information, which improves the naturalness and expressiveness in the synthesized speech.

\textbf{Pre-trained Representation in TTS.}
It is believed that prosody can also be inferred from language information in both current and surrounding utterances \citep{shen2018natural,fang2019towards,xu2021improving,zhou2021dependency}. 
Such information is often entailed in vector representations from a pre-trained LM, such as BERT \citep{devlin2018bert}. Some existing work incorporated BERT embeddings at word or subword-level into autoregressive TTS models~\citep{shen2018natural,fang2019towards}. 
More recent work~\citep{xu2021improving} used the chunked and paired sentence patterns from BERT.
Besides, a relational gated graph network with pre-trained BERT embeddings as node inputs~\citep{zhou2021dependency} was used to extract word-level semantic representations, thus enhancing expressiveness.

\textbf{VAEs in TTS.}
VAEs have been widely adopted in TTS systems to explicit model prosody variation.
The training objective of VAE is to maximise $p_{\theta}(\bm{x})$, the data likelihood parameterised by $\theta$, which can be regarded as the marginalisation w.r.t. the latent vector $\bm{z}$ as shown in Eq. (\ref{eq:likelihood}).
\begin{equation}
    p_{\theta}(\bm{x})=\int p_{\theta}(\bm{x}\mid{\bm{z}}) p(\bm{z}) d \bm{z}.
    \label{eq:likelihood}
\end{equation}
To make this calculation tractable, the marginalisation is approximated using evidence lower bound (ELBO): 
\begin{align}
    \mathcal{L}_\text{ELBO}(\bm{x}) & = 
    \mathbb{E}_{q_{\phi}({\bm{z}} |\bm{x})} 
    [\log p_{\theta}(\bm{x}|\bm{z})] \nonumber\\ &- 
    \beta D_\mathrm{KL}\left(q_{\phi}(\bm{z} | \bm{x}) \| p(\bm{z})\right),
    \label{eq:elbo}
\end{align}

where $q_{\phi}(\bm{z}|\bm{x})$ is the posterior distribution of the latent vector parameterized by $\phi$, $\beta$ is a hyperparameter, and $D_\mathrm{KL}(\cdot)$ is the Kullback-Leibler divergence. The first term measures the expected reconstruction performance of the data from the latent vector and is approximated by Monte Carlo sampling of $\bm{z}$ according to the posterior distribution. The reparameterization trick is applied to make the sampling differentiable. The second term encourages the posterior distribution to approach the prior distribution which is sampled from during inference, and $\beta$ weighs this term's contribution.

A large body of previous work on VAE-based TTS used VAEs to capture and disentangle data variations in different aspects in the latent space.
Works by \citet{akuzawa2018expressive} leveraged VAE to model the speaking style of an utterance. Meanwhile, \citet{hsu2018hierarchical,hsu2018disentangle} explored the disentanglement between prosody variation and speaker information using VAE together with adversarial training. Recently, fine-grained VAE~\citep{sun2020generating,sun2020finegrainedvae} was adopted to model prosody in the latent space for each phoneme or word.
Moreover, vector-quantised VAE was also applied to discrete duration modelling by \citet{yasuda2021end}.

CVAE is a variant of VAE when the data generation is conditioned on some other information $\bm{y}$. In CVAE, both prior and posterior distributions are conditioned on additional variables, and the data likelihood calculation is modified as shown below:
\begin{equation}
    p_{\theta}(\bm{x}\mid\bm{y})=\int p_{\theta}(\bm{x}\mid\bm{z},\bm{y}) p_{\phi}(\bm{z}\mid\bm{y})d\bm{z}.
    \label{eq:cvae_likelihood}
\end{equation}
Similar to VAE, this intractable calculation can be converted to the ELBO form as
\begin{equation}
\begin{aligned}
   \mathcal{L}_{\text{ELBO}}(\bm{x}\mid\bm{y}) & =
    \mathbb{E}_{q_{\phi}(\bm{z} \mid \bm{x},\bm{y})}
    [\log p_{\theta}(\bm{x} \mid \bm{z}, \bm{y})] \nonumber\\ &-
    \beta D_\mathrm{KL}\left(q_{\phi}(\bm{z} \mid \bm{x},\bm{y}) \| p(\bm{z}\mid\bm{y})\right).
\label{eq:cvae_elbo}
\end{aligned}
\end{equation}
To model the conditional prior, a density network is usually used to predict the mean and variance based on the conditional input $\bm{y}$.

\section{CUC-VAE TTS System}
\label{sec: model}

The proposed CUC-VAE TTS system, which is adapted from FastSpeech 2 as shown in Fig.~\ref{fig:model}, aims to synthesize speech with more expressive prosody.
Fig.~\ref{fig:model} describes the model architecture,
which has two components: CU-embedding and CU-enhanced CVAE.
The CUC-VAE TTS system takes as input $[\bm{u}_{i-L},\cdots,\bm{u}_i,\cdots,\bm{u}_{i+L}]$, $\bm{s}_i$ and $\bm{x}_i$, where $[\bm{u}_{i-L},\cdots,\bm{u}_i,\cdots,\bm{u}_{i+L}]$ is the cross-utterance set that includes the current utterance $\bm{u}_i$ and the $L$ utterances before and after $\bm{u}_i$. Each $\bm{u}$ represents the text content of an utterance. Note that $\bm{s}_i$ is the speaker ID, and $\bm{x}_i$ is the reference mel-spectrogram of the current utterance $\bm{u}_i$.
In this section, the two main components of the CUC-VAE TTS system will be introduced in detail.

\subsection{Cross-Utterance Embedding}
The CU-embedding encodes not only the phoneme sequence and speaker information but also cross-utterance information into a sequence of mixture encodings in place of a standard embedding. 
As shown in Fig.~\ref{fig:model}, the first $L$ utterances and the last $L$ utterances surrounding the current one, $\bm{u}_i$, are used as text input in addition to the current utterance and speaker information.
Same as the standard embedding, an extra G2P conversion is first performed to convert the current utterance into phonemes $\bm{P}_i=[p_1,p_2,\cdots,p_T]$, where $T$ is the number of phonemes.
Then, a Transformer encoder is used to encode the phoneme sequence into a sequence of phoneme encodings. Besides, speaker information is encoded into a speaker embedding $\bm{s}_i$ which is directly added to each phoneme encoding to form the mixture encodings $\bm{F}_i$ of the phoneme sequence.
\begin{equation}
\bm{F}_i = [\bm{f}_i(p_1),\bm{f}_i(p_2),\cdots,\bm{f}_i(p_T)],
\end{equation}
where $\bm{f}$ represents resultant vector from the addition of each phoneme encoding and speaker embedding. 

To supplement the text information from the current utterance to generate natural and expressive audio, cross-utterance BERT embeddings together with a multi-head attention layer are used to capture contextual information.
To begin with, $2L$ cross-utterance pairs, denoted as $\bm{C}_i$, are derived from $2L+1$ neighboring utterances $[\bm{u}_{i-L},\cdots,\bm{u}_i,\cdots,\bm{u}_{i+L}]$ as:

\begin{small}
\begin{equation}
\resizebox{\linewidth}{!}{
$
    \bm{C}_i = [c(\bm{u}_{i-L}, \bm{u}_{i-L+1}),\cdots, c(\bm{u}_{i-1}, \bm{u}_{i}),\cdots, c(\bm{u}_{i+L-1}, \bm{u}_{i+L})],
$
    }
    \label{eq:uttpairs}
\end{equation}
\end{small}
where $c(u_k,u_{k+1})= \{[\text{CLS}], \bm{u}_k, [\text{SEP}], \bm{u}_{k+1}\}$, which adds a special token [CLS] at the beginning of each pair and inserts another special token [SEP] at the boundary of each sentence to keep track of BERT.
Then, the $2L$ cross-utterance pairs are fed to the BERT to capture cross-utterance information, which yields $2L$ BERT embedding vectors by taking the output vector at the position of the [CLS] token and projecting each to a 768-dim vector for each cross-utterance pair, as shown below:
$$
\bm{B}_i = [\bm b_{-L}, \bm b_{-L+1}, \cdots, \bm b_{L-1}],
$$
where each vector $\bm b_k$ in $\bm B_i$ represents the BERT embedding of the cross-utterance pair $c(\bm u_k,\bm u_{k+1})$. Next, to extract CU-embedding vectors for each phoneme specifically, a multi-head attention layer is added to combine the $2L$ BERT embeddings into one vector as shown in Eq.~\eqref{eq:mha}.
\begin{equation}
    \bm{G}_{i}=\text {MHA}(\bm{F_i} \bm{W}^{\text{Q}}, \bm{B}_i \bm{W}^{\text{K}}, \bm{B}_i \bm{W}^{\text{V}}),
    \label{eq:mha}
\end{equation}
where MHA$(\cdot)$ denotes the multi-head attention layer, $\bm{W}^{\text{Q}}$, $\bm{W}^{\text{K}}$ and $\bm{W}^{\text{V}}$ are linear projection matrices, and $\bm{F}_i$ denotes the sequence of mixture encodings for the current utterance which acts as the query in the attention mechanism. 
For simplicity, we denote Eq.~\eqref{eq:mha} as $\bm{G}_{i}=[\bm{g}_{1}, \bm{g}_{2},\cdots,\bm{g}_{T}]$ from the multi-head attention being of length $T$ and each of them is then concatenated with its corresponding mixture encoding. 
The concatenated vectors are projected by another linear layer to form the final output $\bm{H}_i$ of the CU-embedding, $\bm{H}_i=[\bm{h}_1,\bm{h}_2,\cdots,\bm{h}_T]$ of the current utterance, as shown in Eq.~\eqref{eq:proj}.
\begin{equation}
    \bm{h}_{t} = [\bm{g}_{t},\bm{f}(p_t)]\bm{W},
    \label{eq:proj}
\end{equation}
where $\bm{W}$ is a linear projection matrix.
Moreover, an additional duration predictor takes $\bm{H}_i$ as inputs and predicts the duration $\bm D_i$ of each phoneme.

\subsection{Cross-Utterance Enhanced CVAE}
In addition to the CU-embedding, a CU-enhanced CVAE is proposed to conquer the lack of prosody variation of FastSpeech 2 and the inconsistency between the standard Gaussian prior distribution sampled by the VAE based TTS system and the true prior distribution of speech.
Specifically, the CU-enhanced CVAE consists of an encoder module and a decoder module, as shown in Fig.~\ref{fig:model}.
The utterance-specific prior in the encoder aims to learn the prior distribution $\bm z_p$ from the CU-embedding output $\bm H$ and predicts duration $\bm D$. 
For convenience, the subscript $i$ is omitted in this subsection.
Furthermore, the posterior module in the encoder takes as input reference mel-spectrogram $\bm x$, then model the approximate posterior $\bm z$ conditioned on utterance-specific conditional prior $\bm z_p$.
Sampling is done from the estimated prior by the utterance-specific prior module and is reparameterized as: 
\begin{equation}
\label{eq:sample_z}
\bm z = \bm{\mu} \oplus \bm{\sigma} \otimes \bm z_p,
\end{equation}
where $\bm \mu$ and $\bm \sigma$ are estimated from conditional posterior module to approximate posterior distribution $\mathcal{N}(\bm \mu,\bm \sigma)$, $\bm z_p$ is sampled from the learned utterance-specific prior, and $\oplus,\otimes$ are elementwise addition and multiplication operation.
Furthermore, the utterance-specific conditional prior module is conducted to learn utterance-specific prior with CU-embedding output $\bm H$ and $\bm D$. 
The reparameterization is as follows:
\begin{equation}
\label{eq:sample_zp}
\bm z_p = \bm \mu_p \oplus \bm \sigma_p \otimes \bm \epsilon,
\end{equation}
where $\bm \mu_p,\bm \sigma_p$ are learned from the utterance-specific prior  module, and $\bm \epsilon$ is sampled from the standard Gaussian $\mathcal{N}({0,1})$.
By substituting Eq.~\eqref{eq:sample_zp} into Eq.~\eqref{eq:sample_z}, the following equation can be derived for the total sampling process:
\begin{equation}
\label{eq:sample_p_new}
\bm z = \bm \mu \oplus \bm \sigma \otimes \bm \mu_p \oplus \bm \sigma \otimes \bm \sigma_p\otimes \bm \epsilon.
\end{equation} 
During inference, sampling is done from the learned utterance-specific conditional prior distribution $\mathcal{N}(\bm \mu_p,\bm \sigma_p)$ from CU-embedding instead of a standard Gaussian distribution $\mathcal{N}({0,1})$.
For simplicity, we can formulate the data likelihood calculation as follows, where the intermediate variable utterance-specific prior $\bm z_p$ from $\bm D,\bm H$ to obtain $\bm z$ is omitted:
\begin{equation}
    \label{eq:our_likehood}
\resizebox{.45\textwidth}{!}{
$
    p_\theta(\bm x \mid \bm{H}, \bm{D})=\int p_\theta(\bm x \mid \bm z, \bm{H}, \bm{D})p_\phi(\bm z\mid \bm{H}, \bm{D}) d \bm z,
$}
\end{equation}
In Eq.~\eqref{eq:our_likehood}, $\phi,\theta$ are the encoder and decoder module parameters of the CUC-VAE TTS system.

Moreover, the decoder in CU-enhanced CVAE is adapted from FastSpeech 2.
An additional projection layer is firstly added to project $\bm z$ to a high dimensional space so that $\bm z$ could be added to $\bm H$.
Next, a length regulator expands the length of input according to the predicted duration $\bm D$ of each phoneme.
The rest of Decoder is same as the Decoder module in FastSpeech 2 to convert the hidden sequence into an mel-spectrogram sequence via parallelized calculation.

Therefore, the ELBO objective of the CUC-VAE can be expressed as,
\begin{equation}
\resizebox{\linewidth}{!}{
$
\begin{aligned}
\mathcal{L}(\bm x\mid{\bm H, \bm{D}})&=\mathbb{E}_{q_{\phi}(\bm{z} \mid{\bm D, \bm H})} [\log p_\theta(\bm x \mid \bm z, \bm D, \bm H)]\\
&-\beta_1 \sum_{n=1}^{T} D_{\mathrm{KL}}\left(q_{\phi_1}\left({\bm z^n} \mid{\bm z_p^n, \bm x}\right) \| q_{\phi_2} \left({\bm z_p^n} \mid{\bm D, \bm H}\right)\right)\\
&-\beta_2 \sum_{n=1}^{T} D_{\mathrm{KL}}\left(q_{\phi_2}\left({\bm z_p^n} \mid{\bm D, \bm H}\right) \| p(\bm z_p^n)\right),
\end{aligned}
$
}
\end{equation}
where $\phi_1,\phi_2$ are two parts of CUC-VAE encoder $\phi$ to obtain $\bm z$ from $\bm z_p, \bm x$ and $\bm z_p$ from $\bm D,\bm H$ respectively, $\beta_1,\beta_2$ are two balance constants, $p(\bm z_p^n)$ is chosen to be standard Gaussian $\mathcal{N}(0,1)$.
Meanwhile, $\bm z^n$ and $\bm z_p^n$ correspond to the latent representation for the $n$-th phoneme, and $T$ is the length of the phoneme sequence.

\section{Experimental Setup}
\label{sec:exp_setup}

\subsection{Dataset} 
To evaluate the proposed CUC-VAE TTS system, a series of experiments were conducted on a single speaker dataset and a multi-speaker dataset.
For the single speaker setting, the LJ-Speech read English data~\citep{LJ-Speech} was used which consists of 13,100 audio clips with a total duration of approximately 24 hours. 
A female native English speaker read all the audio clips, and the scripts were selected from 7 non-fiction books.
For the multi-speaker setting, the train-clean-100 and train-clean-360 subsets of the LibriTTS English audiobook data~\citep{LibriSpeech} were used.
These subsets used here consist of 1151 speakers (553 female speakers and 598 male speakers) and about 245 hours of audio.
All audio clips were re-sampled at 22.05 kHz in experiments for consistency.

The proposed CU-embedding in our system learns the cross-utterance representation from surrounding utterances.
However, unlike LJ-Speech, transcripts of LibriTTS utterances are not arranged as continuous chunks of text in their corresponding book.
Therefore, transcripts of the LibriTTS dataset were pre-processed to find the location of each utterance in the book, so that the first $L$ and last $L$ utterances of the current one can be efficiently obtained during training and inference.
The pre-processed scripts and our code are available
\renewcommand{\thefootnote}{\arabic{footnote}}
\footnote{\url{https://github.com/NeuroWave-ai/CUCVAE-TTS}}.

\subsection{System Specification} 
The proposed CUC-VAE TTS system was based on the framework of FastSpeech 2.
The CU-embedding utilised a Transformer to learn the current utterance representation, where the dimension of phoneme embeddings and the size of the self-attention were set to 256. 
To explicitly extract speaker information, 256-dim speaker embeddings were also added to the Transformer output.
Meanwhile, the pre-trained BERT model to extract cross-utterance information had 12 Transformer blocks and 12-head attention layers with 110 million parameters. The size of the derived embeddings of each cross-utterance pair was 768-dim. Note that the BERT model and corresponding embeddings were fixed when training the TTS system.
Network in CU-enhanced CVAE consisted of four 1D-convolutional (1D-Conv) layers with kernel sizes of 1 to predict the mean and variance of 2-dim latent features.
Then a linear layer was added to transform the sampled latent feature to a 256-dim vector.
The duration predictor which consisted of two convolutional blocks and an extra linear layer to predict the duration of each phoneme for the length regulator in FastSpeech 2 was adapted to take in CU-embedding outputs.
Each convolutional block was comprised of a 1D-Conv network with ReLU activation followed by a layer normalization and dropout layer.
The Decoder adopted four feed-forward Transformer blocks to convert hidden sequences into 80-dim mel-spectrogram sequence, similar to FastSpeech 2.
Finally, HifiGAN~\citep{hifigan} was used to synthesize waveform from the predicted mel-spectrogram. 

\subsection{Evaluation Metrics}
In order to evaluate the performance of our proposed component, both subjective and objective tests were performed.
First of all, a subjective listening test was performed over 11 synthesized audios with 23 volunteers asked to rate the naturalness of speech samples on a 5-scale mean opinion score (MOS) evaluation. The MOS results were reported with $95\%$ confidence intervals. 
In addition, an AB test was conducted to compare the CU-enhanced CVAE with utterance-specific prior and normal CVAE with standard Gaussian prior.
23 volunteers were asked to choose the preference audio generated by different models in the AB test.

For the objective evaluation, $F_0$ frame error (FFE)~\citep{FFE} and mel-cepstral distortion (MCD)~\citep{MCD} were used to measure the reconstruction performance of different VAEs.
FFE combined the Gross Pitch Error (GPE) and the Voicing Decision Error (VDE) and was used to evaluate the reconstruction of the $F_0$ track.
MCD evaluated the timbral distortion, which was computed from the first 13 MFCCs in our experiments.
Moreover, word error rates (WER) from an ASR model trained on the real speech from the LibriTTS training set were reported. 
Complementary to naturalness, the WER metric showed both the intelligibility and the degree of inconsistency between synthetic speech and real speech. The ASR system used in this paper was an attention-based encoder-decoder model trained on Librispeech 960-hour data, with a WER of 4.4\% on the test-clean set.
Finally, the diversity of samples was evaluated by measuring the standard deviation of two prosody attributes of each phoneme: relative energy ($E$) and fundamental frequency
($F_0$), similar to \citet{sun2020finegrainedvae}. Relative energy was calculated as the ratio of the average signal amplitude within a phoneme to the average amplitude of the entire sentence, and fundamental frequency was measured using a pitch tracker.
In this paper, the average standard deviation of $E$ and $F_0$ of three phonemes in randomly selected 11 utterances was reported to evaluate the diversity of generated speech.

\section{Results}
\label{sec:exp_results}
This section presents the series of experiments for the proposed CUC-VAE TTS system.
First, ablation studies were performed to progressively show the influence of different parts in the CUC-VAE TTS system based on MOS and WER.
Next, the reconstruction performance of CUC-VAE was evaluated by FFE and MCD.
Then, the naturalness and prosody diversity using CUC-VAE were compared to FastSpeech 2 and other VAE techniques.
At last, a case study illustrated the prosody variations with different cross-utterance information as an example. 
The audio examples are available on the demo page
\renewcommand{\thefootnote}{\arabic{footnote}}
\footnote{ \url{http://bit.ly/cuc-vae-tts-demo}}.

\subsection{Ablation Studies}
Ablation studies in this section were conducted on the LJ-Speech data based on the subjective test and WER.
First, to investigate the effect of the different number of neighbouring utterances, CUC-VAE TTS systems built with $L=1,3,5$ were evaluated using MOS scores, as shown in Table~\ref{tab:ablation_study2}.

\begin{table}[ht]
    \centering
    \caption{The MOS results of CUC-VAE TTS systems on LJ-Speech dataset. MOS was reported with $95\%$ confident intervals. ``$L=1$'',``$L=3$'',``$L=5$'' represented the number of past and future utterances.}
    \resizebox{\linewidth}{!}{
    \begin{tabular}{ccc}
    \toprule[1pt]
     Systems &
     Cross-utterance ($2L$) &
    \textbf{MOS}\\
    \hline
         CUC-VAE &$L=1$& 2.93 $\pm$ 0.12\\
         CUC-VAE &$L=3$& 3.72 $\pm$ 0.09\\
         CUC-VAE &$L=5$& \textbf{3.95 $\pm$ 0.07}\\
    \bottomrule[1pt]
    \end{tabular}
    }
    \label{tab:ablation_study2}
\end{table}
The effect of the different number of neighbouring utterances on the naturalness of the synthesized speech can be observed by comparing MOS scores which is the higher the better. 
The CUC-VAE with $L=5$ achieved highest score 3.95 compared to system with $L=1$ and $L=3$.
Since only marginal MOS improvements were obtained using more than $5$ neighbouring utterances, the rest of experiments were performed using $L=5$. 

Then we investigated the influence of each part of CUC-VAE on performance.
The baseline was our implementation of Fastspeech 2.
For the system denoted as Baseline + fine-grained VAE which served as a stronger baseline, the pitch predictor and energy predictor of FastSpeech 2 were replaced with a fine-grained VAE with 2-dim latent space.
Based on the fine-grained VAE baseline, the CVAE was added without the CU-embedding to the system, referred to as Baseline+CVAE to verify the function of CVAE on the system, which conditions on the current utterance. Again, MOS was compared among these systems as shown in Table \ref{tab:ablation_study1}.

\begin{table}[ht]
    \centering
    \caption{The MOS results of TTS systems with different modules on LJ-Speech dataset. MOS was reported with $95\%$ confident intervals. Baseline + fine-grained VAE added a fine-grained VAE to baseline. Baseline+CVAE represents a CVAE TTS system without CU-embedding. }
    \resizebox{0.8\linewidth}{!}{
    \begin{tabular}{cc}
    \toprule[1pt]
     Systems &
    \textbf{MOS}\\
    \hline
        Ground Truth& 4.31 $\pm$ 0.06\\
        \hdashline
         Baseline& 3.85 $\pm$ 0.07\\
         Baseline+Fine-grained VAE& 3.55 $\pm$ 0.08\\
         Baseline+CVAE& 3.64 $\pm$ 0.08\\
         CUC-VAE& \textbf{3.95 $\pm$ 0.07}\\
    \bottomrule[1pt]
    \end{tabular}
    }
    \label{tab:ablation_study1}
\end{table}

As shown in Table~\ref{tab:ablation_study1}, MOS progressively increased when fine-grained VAE, CVAE, and CU-embedding were added in consecutively. The proposed CUC-VAE TTS system achieved the highest MOS 3.95 compared to baselines.
The results indicated that CUC-VAE module played a crucial role in generating more natural audio. 

To verify the importance of the utterance-specific prior to the synthesized audio, the same CUC-VAE system was used, and the only difference is whether to sample latent prosody features from the utterance-specific prior or from a standard Gaussian distribution. 
A subjective AB test was performed which required 23 volunteers to provide their preference between audios synthesized from the 2 approaches. Moreover, WER was also compared here to show the intelligibility of the synthesized audio.
As shown in Table~\ref{tab:ABtest}, the preference rate of using the utterance-specific prior is 0.52 higher than its counterpart, and a 4.9\% absolute WER reduction was found, which confirmed the importance of the utterance-specific prior in our CUC-VAE TTS system.

\begin{table}[ht]
    \centering
    \caption{The subjective listening preference rate between CUC-VAE with or without utterance-specific prior from the AB test. The CUC-VAE without utterance-specific prior was a simplified version of our proposed CUC-VAE where latent samples were drawn from a standard Gaussian distribution instead of utterance-specific prior. WER metric was also reported.}
    \resizebox{\linewidth}{!}{
    \begin{tabular}{cccc}
    \toprule[1pt]
     System &utterance-specific prior &
    \textbf{RATE}&
    \textbf{WER}\\
    \hline
        CUC-VAE & \XSolidBrush  & 0.24 & 14.8\\
        CUC-VAE &\Checkmark& \textbf{0.76} & \textbf{9.9} \\
    \bottomrule[1pt]
    \end{tabular}
    }
    \label{tab:ABtest}
\end{table}

\subsection{Reconstruction Performance}
FFE and MCD were used to measure the reconstruction performance of VAE systems. An utterance-level prosody modelling baseline which extract one latent prosody feature vector for an utterance was added for more comprehensive comparison, and is referred to as the Global VAE.
\begin{table}[htbp]
    \centering
    \caption{Reconstruction preformance on LJ-Speech and LibriTTS dataset. + Global VAE and + fine-grained VAE represent that the baseline is added the global VAE and the fine-grained VAE, respectively.}
     \resizebox{\linewidth}{!}{
    \begin{tabular}{ccccc}
    \toprule[1pt]
    \multirow{2}{*}{Systems} &
    \multicolumn{2}{c}{\textbf{LJ-Speech}}&
    \multicolumn{2}{c}{\textbf{LibriTTS}}\\
    \cmidrule(lr){2-3} \cmidrule(lr){4-5}
    &\textbf{MCD}&
    \textbf{FFE}&
    \textbf{MCD}&
    \textbf{FFE}\\
    \hline
         Baseline&6.70&0.58 & 6.32 & 0.58\\
         Baseline+Global VAE&6.50&0.41 & 6.27 &0.45\\
         Baseline+Fine-grained VAE&6.34&0.26 & 6.28 & 0.35\\ 
         CUC-VAE& \textbf{6.27}&\textbf{0.24}&\textbf{6.04} &\textbf{0.34}\\
    \bottomrule[1pt]
    \end{tabular}
    }
    \label{tab:rp}
\end{table}

\begin{table*}[ht]
    \centering
    \caption{Sample naturalness and diversity results on LJ-Speech and LibriTTS datasets.
    Three metrics are reported for each dataset, namely MOS, WER, and Prosody Std.
    The Prosody Std. includes standard deviations of relative energy ($E$) and fundamental frequency ($F_0$) in Hertz within each phonene.}
    \resizebox{\textwidth}{!}{
    \begin{tabular}{ccccccccc}
    \toprule[1pt]
    \multicolumn{1}{c}{}&
    \multicolumn{4}{c}{\textbf{LJ-Speech}} &
    \multicolumn{4}{c}{\textbf{LibriTTS}} \\
    \cmidrule(lr){2-5} \cmidrule(lr){6-9}
    \multirow{2}{*}{}&
    \multirow{2}{*}{\textbf{MOS}}&
    \multirow{2}{*}{\textbf{WER}}&
    \multicolumn{2}{c}{\textbf{Prosody Std.}}&
    \multirow{2}{*}{\textbf{MOS}}&
    \multirow{2}{*}{\textbf{WER}}&
    \multicolumn{2}{c}{\textbf{Prosody Std.}}\\
 \cmidrule(lr){4-5} \cmidrule(lr){8-9}
  &&& $\bm{F}_0$ & $\bm{E}$ &&& $\bm{F}_0$ & $\bm{E}$  \\   
    \hline
        Ground Truth& 4.31 $\pm$ 0.06 & 8.8 &- &-& 4.10 $\pm$ 0.07 & 5.0 & - & -\\
         Baseline&3.85 $\pm$ 0.07 &10.8&$1.86 \times 10^{-13}$&$6.78 \times 10^{-7}$&3.53 $\pm$ 0.08&6.0&$2.13\times 10^{-13}$ &$7.22\times 10^{-7}$\\
         Baseline+Global VAE&3.82 $\pm$ 0.07 &10.4&1.46& 0.0004&3.59 $\pm$ 0.08&10.8&2.01 & 0.0054 \\
         Baseline+Fine-grained VAE&3.55 $\pm$ 0.08 &12.8&49.60 & 0.0670&3.43 $\pm$ 0.08& 5.6 & 63.64 & 0.0901\\ 
         CUC-VAE&\textbf{3.95} $\bm\pm$ \textbf{0.07}&\textbf{9.9}& 26.35 & 0.0184&\textbf{3.63} $\bm\pm$ \textbf{0.08} &\textbf{5.5} & 30.28& 0.0217\\
    \bottomrule[1pt]
    \end{tabular}
    }
    \label{tab:nd}
\end{table*}

Table.~\ref{tab:rp} shows the reconstruction performance on the LJ-Speech dataset and LibriTTS dataset, respectively.
Baseline had the highest value of FFE and MCD on the LJ-Speech dataset and LibriTTS dataset.
The value of FFE and MCD decreased when the global VAE was added and was further reduced when the fine-grained VAE was added to the baseline.
Our proposed CUC-VAE TTS system achieved the lowest FFE and MCD across the table on both the LJ-Speech and LibriTTS datasets.
This indicated that richer prosody-related information entailed in both cross-utterance and conditional inputs was captured by CUC-VAE.

\subsection{Sample Naturalness and Diversity}
Next, sample naturalness and intelligibility were measured using MOS and WER respectively on both LJ-Speech and LibriTTS datasets. 
Complementary to the naturalness, the diversity of generated speech from the conditional prior was evaluated by comparing the standard deviation of $E$ and $F_0$ similar to \cite{sun2020finegrainedvae}.

LJ-Speech experiments were shown in left part of Table.~\ref{tab:nd}. 
Compared to the global VAE and fine-grained VAE, the proposed CUC-VAE received the highest MOS and achieved the lowest WER.
Although both $F_0$ and $E$ of the CUC-VAE TTS system were lower than the baseline + fine-grained VAE, the proposed system achieved a clearly higher prosody diversity than the baseline and baseline + global VAE systems.
The fine-grained VAE achieved the highest prosody variation as its latent prosody features were sampled from a standard Gaussian distribution, which lacks the constraint of language information from both the current and the neighbouring utterances. This caused extreme prosody variations to occur which impaired both the naturalness and the intelligibility of synthesized audios.
As a result, the CUC-VAE TTS system was able to achieve high prosody diversity without hurting the naturalness of the generated speech. In fact, the adequate increase in prosody diversity improved the expressiveness of the synthesized audio, and hence increased the naturalness.

The right part of Table.~\ref{tab:nd} showed the results on LibriTTS dataset. Similar to the LJ-Speech experiments, the CUC-VAE TTS system achieved the best naturalness measured by MOS, the best intelligibility measured by WER, and the second-highest prosody diversity across the table.
Overall, consistent improvements in both naturalness and prosody diversity were observed on both single-speaker and multi-speaker datasets.

\subsection{A Case Study}
\begin{figure}
\centering
\subfigure[Who asked the time? \textbf{Mary asked the time.}]{\label{fig:subfig:a}
\includegraphics[width=\linewidth]{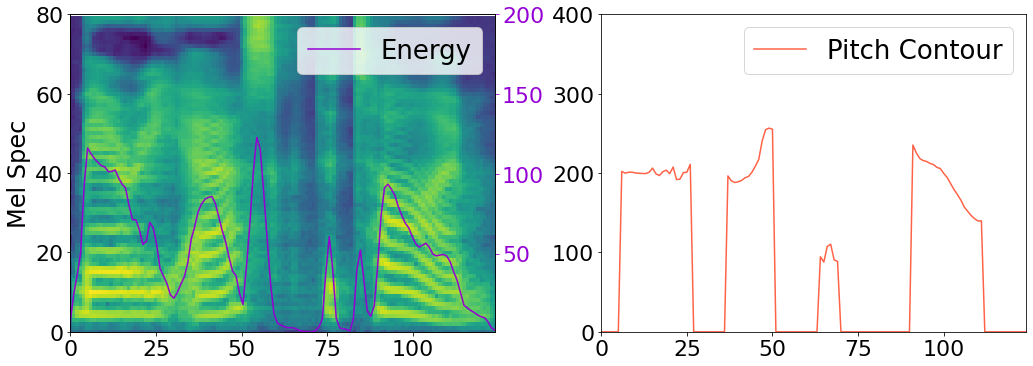}}
\vfill
\subfigure[\textbf{Mary asked the time,} and was told it was only five.]{\label{fig:subfig:b}
\includegraphics[width=\linewidth]{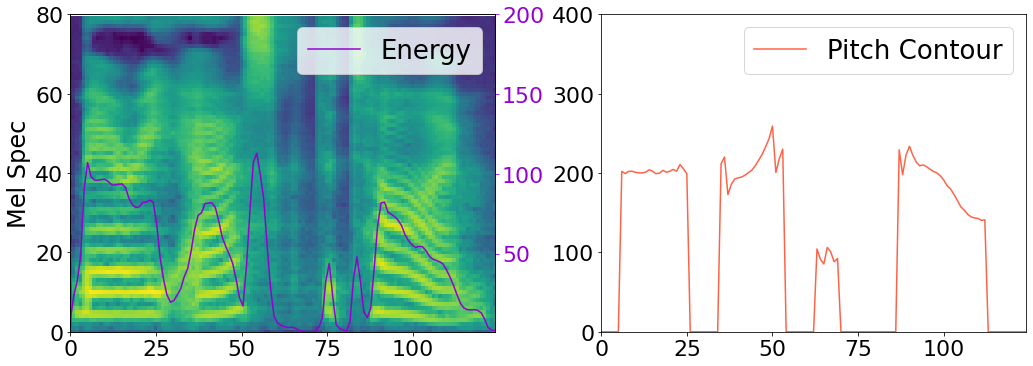}}
\caption{Comparisons between the energy and pitch contour of same text “Mary asked the time" but different neighbouring utterances, generated by CUC-VAE TTS trained on LJ-Speech.}
\label{fig:cs}
\end{figure}

To better illustrate how the utterance-specific prior influenced the naturalness of the synthesized speech under a given context, a case study was performed by synthesizing an example utterance, “Mary asked the time”, with two different neighbouring utterances: “Who asked the time? Mary asked the time.” and “Mary asked the time, and was told it was only five.” Based on the linguistic knowledge, to answer the question in the first setting, an emphasis should be put on the word “Mary”, while in the second setting, the focus of the sentence is “asked the time”. The model trained on LJ-Speech dataset was used to synthesize the utterance and the results were shown in Fig.~\ref{fig:cs}. 

Fig.~\ref{fig:cs} showed the energy and pitch of the two utterance. Energy of the first word “Mary” in Fig.~\ref{fig:subfig:a} changed significantly (energy of “Ma-” was much higher than “-ry”), which reflected an emphasis on the word “Mary”, whereas in Fig.~\ref{fig:subfig:b}, energy of “Mary” had no obvious change, i.e., the word was not emphasized. 
On the other hand, the fundamental frequency of words “asked” and “time” stayed at a high level for a longer time in the second audio than the first one, reflecting another type of emphasis on those words which was also coherent with the given context. Therefore, the difference of energy and pitch between the two utterances demonstrated that the speech synthesized by our model is sufficiently contextualized.

\section{Conclusion}
\label{sec: con}
In this paper, a non-autoregressive CUC-VAE TTS system was proposed to synthesize speech with better naturalness and more prosody diversity.
CUC-VAE TTS system estimated the posterior distribution of latent prosody features for each phone based on cross-utterance information in addition to the acoustic features and speaker information. 
The generated audio was sampled from an utterance-specific prior distribution, approximated based on cross-utterance information.
Experiments were conducted to evaluate the proposed CUC-VAE TTS system with metrics including MOS, preference rate, WER, and the standard deviation of prosody attributes.
Experiment results showed that the proposed CUC-VAE TTS system improved both the naturalness and prosody diversity in the generated audio samples, which outperformed the baseline in all metrics with clear margins.

\bibliography{custom}
\bibliographystyle{acl_natbib}




\end{document}